\documentclass{mem}
\usepackage{natbib}\usepackage{txfonts}\usepackage{balance}
\usepackage{graphicx}
\usepackage[breaklinks]{hyperref}
\usepackage{etoolbox}
\makeatletter
\patchcmd\@combinedblfloats{\box\@outputbox}{\unvbox\@outputbox}{}{\errmessage{\noexpand patch failed}}
\makeatother
\pdfoutput=1
\begin{document}
\def\teff{$T\rm_{eff }$}
\def\kms{$\mathrm {km s}^{-1}$}
\newcommand{\eprint}{}

\bibpunct{(}{)}{;}{a}{}{,} 


\title{
Magnetic Cataclysmic Variables discovered in hard X-rays
}

   \subtitle{}

\author{
M. Falanga\inst{1},
D. de Martino\inst{2},
F. Bernardini\inst{3,2},
\and K. Mukai\inst{4,5}
          }

\institute{
International Space Science Institute (ISSI), Hallerstrasse 6, CH-3012 Bern, Switzerland
\and INAF - Osservatorio Astronomico di Capodimonte, Salita Moiariello 16, I-80131 Napoli, Italy
\and INAF - Osservatorio Astronomico di Roma, via Frascati 33, Monteporzio Catone, I-00040 Roma, Italy
\and CRESST and X-Ray Astrophysics Laboratory, NASA Goddard Space Flight Center, Greenbelt, MD 20771, USA
\and Department of Physics, University of Maryland, Baltimore County, 1000 Hilltop Circle, Baltimore, MD 21250, USA\\
\email{mfalanga@issibern.ch}
}

\authorrunning{Falanga et al.}

\titlerunning{Hard X-ray Cataclysmic Variables}

\abstract{
Among hard X-ray galactic sources detected by {\it INTEGRAL} and {\it Swift} surveys, those discovered as accreting white dwarfs have surprisingly boosted in number, representing 20\% of the galactic sample. The majority are identified as magnetic cataclysmic variabiles of the intermediate polar type suggesting this subclass as an important constituent of galactic population of X-ray sources. In this conference-proceeding, we review the X-ray emission properties as observed with our ongoing {\it XMM-Newton} programme of newly discovered {\it INTEGRAL} and/or {\it Swift} sources that enlarged almost by a factor of two, identifying cataclysmic variabiles commonalities and outliers.
}
\maketitle{}

\section{Introduction}

Cataclysmic variables (CVs) are close binary systems composed by a white dwarf (WD) that accretes matter from a Roche-lobe filling, late-type companion star. Although CVs can be classified into several groups, depending on their observational characteristics, we can distinguish between two main classes: magnetic and non-magnetic CVs. In magnetic cataclysmic variables (mCVs) the accretion of matter onto the compact object is dominated by the magnetic field of the WD. Depending of the field strength, $B$, mCVs can be further divided into two types: polar mCVs have $B\sim10^{7}-10^{8}$~G and, in these systems, the matter outflow from the companion star is immediately funnelled along the field lines and accreted onto the WD polar caps, hence no accretion disk is formed. The strong field 
also causes the WD spin rate to be locked with the binary 
orbital period, i.e., syncronismus (P$_{\rm \omega =\Omega}\sim$hrs). On the contrary, intermediate polar mCVs 
are believed to harbour weakly magnetized  accreting WDs 
with $B \lesssim 5\times 10^{6}$~G. This allows for a fast rotation 
of the WD (i.e., asynchronous WD rotation, P$_{\rm \omega}\sim$mins) and for the formation of an accretion disk. The latter is disrupted at the magnetospheric radius, due to the strong WD magnetic pressure. 

From that point, the matter attaches to the magnetic field lines and follows them almost radially at free-fall velocity towards the WD magnetic poles surface \citep[see e.g.,][]{Aizu1973}. Magnetic accretion produces a strong shock above the WD magnetic poles, so-called post-shock region (PSR), hard optically thin emission, where temperatures can reach $kT_{\rm brem.} \sim 20-40$ keV,  below the flow cools by bremsstrahlung (hard X-rays) and cyclotron (optical) radiation that are partially thermalized and re-emitted in the soft X-rays and/or EUV/UV domains.  The efficiency of the two cooling mechanisms depends on the magnetic field strength: cyclotron is increasingly  efficient in high field systems ($>10$MG, the Polars) and is able to suppress high temperatures. It is then likely that the low-field systems (IPs) preferentially emit in the hard X-rays, being bremsstrahlung dominated, as indeed observed. This actually explains, why we were used to observe the CVs as bright soft X-rays sources during the ROSAT era \citep{Beuermann1995}. 
We note, that among 13 hard Polars 
identified so far \citep[see][Bernardini 2019b, in prep]{Bernardini2014,Bernardini2017,Gabdeev2017,Mukai2017}. Those few with determined magnetic fields (up to $40\times10^{6}$~G) challenge our knowledge of  emission properties in mCVs. Since three hard X-ray Polars are also slightly desynchronised, asynchronism seemed a common characteristics of hard mCVs. Other processes, such as Compton scattering, turn out to be much less important for low mass accretion rates or low WD masses \citep{Suleimanov2008}. Substantial reviews on CVs can be found by \citet[][]{Warner1995,Norton2004,Ferrario2015,Mukai2017}, and references therein.

\begin{figure}[t]
\begin{center}
\vspace{-0.2cm}
{\includegraphics[width=5.5cm, angle=-90]{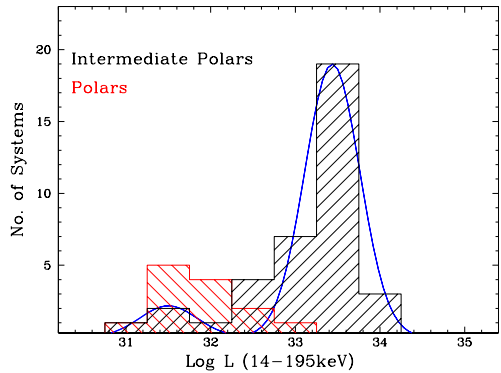}}
\vspace{-0.2cm}
\caption{
\footnotesize
The 14--195\,keV luminosity distribution of confirmed IPs and polars in 
the {\it Swift}/BAT catalogue with {\it Gaia} distance accurate better than
10$\%$. A bimodal distribution (blue line) in the IP sample is suggested with 4 
low-luminosity systems overlapping the polar sample (in red). Figure provided by D. De Martino
}
\label{fig1}
\end{center}
\end{figure}

\section{The hard X-ray Cataclysmic Variables surveys}
\label{sec:distribution}

Our  view of the hard X-ray sky  dramatically changed thanks
to the deep {\it INTEGRAL}/ISGRI and {\it Swift}/BAT surveys with more than 1000 sources detected above 20 keV   
\citep{Bird2016,Krivonos2010,Cusumano2010,Baumgartner2013,Oh2018}.  
These surveys have surprisingly shown a large number of CVs, the majority, $\sim70\%$, being magnetic of the intermediate polar type. mCVs, in particular the IPs, are claimed as important contributors
to the galactic X-ray source population above $\sim
10^{31}\rm erg\,s^{-1}$ from surveys of the galactic centre with  
{\it Chandra}, {\it XMM-Newton}, and {\it NuSTAR} \citep{Muno2004,Heard2013,Perez2015,Hailey2016}, respectively.

Whether IPs  also dominate the galactic ridge 
emission is still  disputed \citep[see e.g.,][]{Revnivtsev2009}. Our knowledge 
of X-ray binary populations is also crucial to understand close binary evolution.
The high ($\sim20-25\%$) incidence of magnetism in CVs with respect to that
in single WDs ($\sim10\%$) \citep[][]{Ferrario2015},
would either imply CV formation is favoured by magnetism or CV production
enhances magnetism \citep{Tout2008}.

The negligible absorption in the hard X-rays makes these 
surveys unique for population studies. In particular the {\it INTEGRAL}/ISGRI and specially
{\it Swift}/BAT survey, with a more uniform exposure over the sky was used to estimate the mCV
space densities but with large uncertainties \citep[][and references therein]{Pretorius2014}. 
The hard X-ray BAT and ISGRI catalogues surveys, however, still carry tentative new CV
identifications, with many claimed as magnetic, based
on optical follow-ups and thus subject of revisions. However, 
the magnetic nature can only be inferred
through the detection or non-detection of Xray
pulses at the WD spin period that imply magnetically
channeled accretion and through the study of broad-band spectra.

The {\it Gaia} DR2 release in April 2018 now offers the opportunity of 
assessing the true space densities. Using the shallow flux-limits 
of the  70-month{\it Swift}/BAT sample of \citet[][]{Pretorius2014} and the {\it Gaia} 
parallaxes, the IP space density appears to be lower than previously estimated
$\rm < 1.3\times10^{-7}\,pc^{-3}$ \citet{Schwope2018}.
The release of the 105-month {\it Swift}/BAT catalogue  reaching flux levels down to  
$\sim 7$ and $\rm 8 \times 10^{-12}\rm erg\,cm^{-2}\,s^{-1}$ 
over 50$\%$ and 90$\%$ of the sky gives the opportunity to confirm this finding 
and to unveil a putative still-hidden low-luminosity IP population from 
hints of bimodality in the luminosity distribution (see Fig. \ref{fig1}).

\section{The spin-orbit period plane of confirmed IPs}
\label{sec:spin-orbit}

\begin{figure}[t]
\vspace{-1.4cm}
{\includegraphics[width=7.0cm]{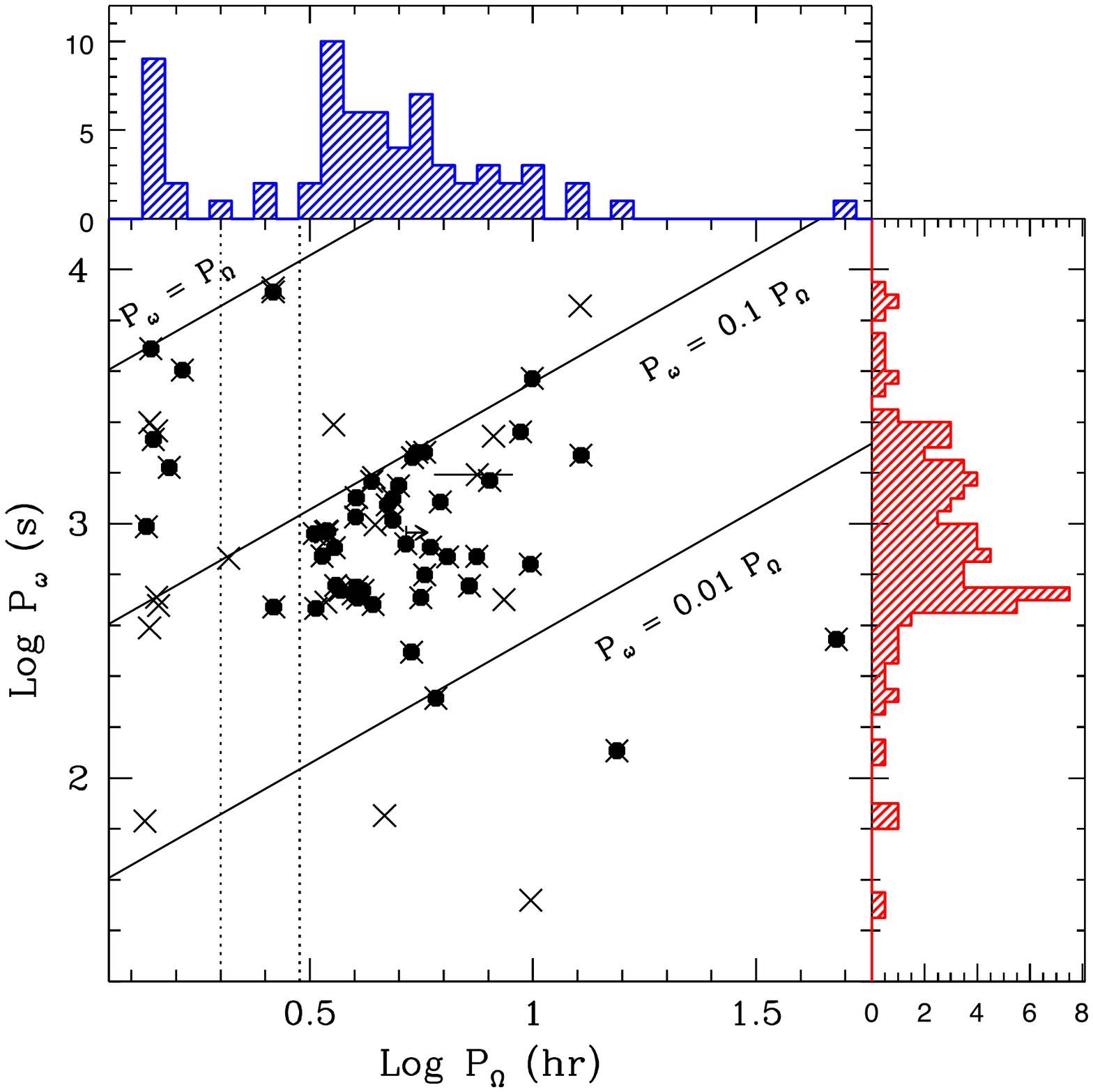}}
\vspace{-2.10cm}
\caption{
\footnotesize
The spin-orbit period plane of confirmed IPs (crosses). Hard X-ray detected
sources are also shown as filled circles. Solid lines
mark synchronism and two levels of asynchronism (0.1 and 0.01). Vertical
lines mark the orbital CV gap, where mass transfer is expected to stop at the upper bound
and to be resumed at the lower bound. The spin (right
panel) and orbital (upper panel) period  distributions are reported as from
\citep{Bernardini2017,Bernardini2018,Bernardini2019}.
}
\label{fig2}
\end{figure}

One of the most challenging open question in the context of CVs is the lack of observations of systems with periods between  $\sim2$ and $\sim3$ hours, known as the period gap. The orbital evolution of CVs with periods shorter than those in the gap is dominated by gravitational radiation while for periods exceeding those of the gap it is dominated by magnetic braking of the secondary star. The  magnetic complexity might then explain the period gap, as well as different population. The low-luminosity IP population are found below the 2--3h CV orbital period gap and thus
they are low-rate accretors (see Fig. \ref{fig2}).  The number of IPs below the 2--3\,h CV orbital period gap 
has surprisingly increased to 10 members. This is challenging since short period mCVs should have already reached synchronism \citet{Norton2008}. The large spread in spin-to-orbit
period ratios of short period IPs may suggest they represent a different population of old, possibly, low-field systems. Clearly the mechanisms driving mCV evolution are still to be understood.

\section{The mass distribution of hard X-ray IPs} 
\label{sec:mass}

The majority of all stars born in the Galaxy will one day evolve into WDs. WDs are supported against collapse by electron degeneracy pressure and as such show the remarkable property that the more massive they are, the smaller their radius. Moreover, this mass-radius relationship sets an upper limit to the mass of a WD \citep{Chandrasekhar1931} above which electron degeneracy can no longer support them, a result that underpins our understanding of type Ia supernovae and hence the expansion of the Universe. The mass-radius relationship forms an essential part of many studies of WD, such as the initial-final mass relationship, stellar evolution, WD population i.e., mass measurements for white dwarfs in interacting binary systems or isolated WDs, WD mergers masses cousing gravitational waves, WD equation of states, i.e., matter composition, the WD luminosity function etc... Despite its huge importance to a wide range of astrophysical topics, the WD mass-radius relationship remains poorly tested observationally. 

\begin{figure}[t]
\vspace{-1.4cm}
{\includegraphics[width=7.0cm]{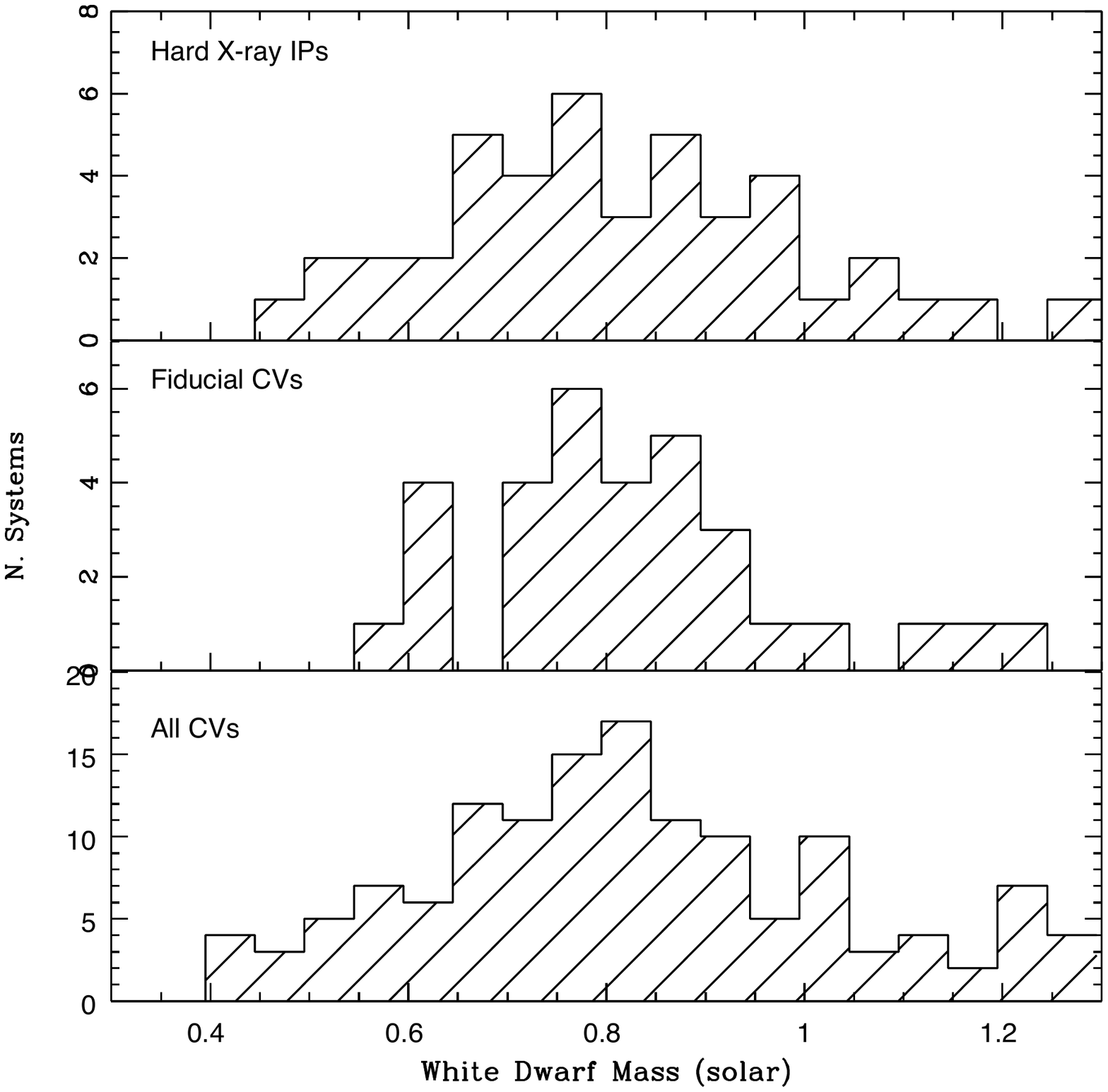}}
\vspace{-2.10cm}
\caption{
\footnotesize
The mass distribution of hard X-ray IPs (top) as derived by us
and by \citet{Brunschweiger2009}, giving
$\rm <M>=0.80\pm0.16\,M_{\odot}$, compared with those of
 "fiducial" CVs \citep{Zorotovic2011}, giving $\rm
<M>=0.82\pm 0.15\,M_{\odot}$ (centre) and  those of CVs
from the Ritter \& Kolb, 7.20v CV Catalogue (bottom), giving $\rm
<M>=0.82\pm0.24\,M_{\odot}$. Figure provided by D. De Martino.
}
\label{fig3}
\end{figure}

The fact, that IPs are mainly discovered in the {\it INTEGRAL}/{\it Swift} hard X-ray broadband energy band, allow us to estimate the IPs masses by measuring the maximum temperature of the post-shock plasma \citep[see e.g.,][]{Suleimanov2005,Falanga2005,deMartino2008,Anzolin2009,Bernardini2012,Bernardini2013,Bernardini2015,Bernardini2017,Bernardini2018,Bernardini2019}, (see Fig. \ref{fig3}). We note, the IP radius can be calculated from the \citet{Nauenberg1972} WD mass-radius relation. 
Using mass determinations from our combined {\it XMM-Newton/INTEGRAL/Swift} 
observations and those from \citet{Brunschweiger2009, Suleimanov2005,Tomsick2016, Shaw2018} 
very massive WDs are not favoured although the majority is found above 0.7\,M$_{\odot}$ 
(Fig.\,3). Other parameters as the local accretion rate and the effects of reflection
are to be included. However, IP masses do not appear much different from other CVs. The mean WD mass among CVs is $\langle M_{\rm WD} \rangle = 0.83 \pm0.19$ M$_{\odot}$, much larger than that found for pre-CVs, $\langle M_{\rm WD} \rangle = 0.67 \pm 0.21$ M$_{\odot}$. It may indicate, that either most CVs have formed above the orbital-period gap (which requires a high WD mass to initiate stable mass transfer or a previous phase of thermal-timescale mass transfer), or the mass of the WDs in CVs grows through accretion (which strongly disagrees with the predictions of classical nova models). Both options may imply that CVs contribute to the single-degenerate progenitors of type Ia supernovae.

\section{Conclusions}

The number of hard X-ray emitting CVs has been boosted in 
the recent years, unveiling the dominance  of asynchronously rotating, 
magnetic accreting white dwarf primaries - the Intermediate Polars. These systems
are disputed to be important contributors to the galactic population of 
low-luminosity X-ray sources. We have been carrying out a systematic identification
programme of new optically discovered hard  X-ray CV candidates with the
unique potential of {\it XMM-Newton} to 
securely assess their purported magnetic nature by detecting X-ray spin pulses and
characterizing  their spectral properties. We aim at obtaining a 
CV flux-limited sample with {\it Gaia} distances finally making 
statistical  studies possible and  uncover the true population of low 
luminosity sources. Up to date 70 confirmed IPs are known, 26 of them identified by us \citep[see e.g., ][]{Bonnet-Bidaud2007,deMartino2008,Anzolin2009,Bernardini2012,Bernardini2013,Bernardini2015,Bernardini2017,Bernardini2018,Bernardini2019}.  The polar group,  instead, amounts  
to $\sim$130 systems with 13 identified as hard X-ray sources, 3 of them
found by us \citep[see e.g.,][2019b in prep.]{Bernardini2014,Bernardini2017}.
suggesting that hard Polars are not as rare as previously thought.
We also disproved the magnetic nature for 6 candidates 
 \citep[see e.g.,][]{Bernardini2014,deMartino2010,deMartino2013}. 
Therefore, secure identifications are needed to construct true  samples.

\begin{acknowledgements}
DdM acknowledges financial suport from INAF-ASI agreements 
I/037/12/0 and ASI-INAF n.2017-14-H.0 and INAF-PRIN SKA/CTA Presidential
Decree 70/2016.
FB is founded by the European Union's 
Horizon 2020 research and innovation programme under the Marie 
Sklodowska-Curie grant agreement n. 664931.

\end{acknowledgements}

\bibliographystyle{aa}
\bibliography{bibl_new}

\end{document}